\documentclass{jfm}
\usepackage{t1enc,alltt,amssymb,amsmath,natbib}
\usepackage{graphicx}

\usepackage{graphicx}
\usepackage{natbib}

\ifCUPmtlplainloaded \else
  \checkfont{eurm10}
  \iffontfound
    \IfFileExists{upmath.sty}
      {\typeout{^^JFound AMS Euler Roman fonts on the system,
                   using the 'upmath' package.^^J}%
       \usepackage{upmath}}
      {\typeout{^^JFound AMS Euler Roman fonts on the system, but you
                   dont seem to have the}%
       \typeout{'upmath' package installed. JFM.cls can take advantage
                 of these fonts,^^Jif you use 'upmath' package.^^J}%
      }
  \else
  \fi
\fi

\ifCUPmtlplainloaded \else
  \checkfont{msam10}
  \iffontfound
    \IfFileExists{amssymb.sty}
      {\typeout{^^JFound AMS Symbol fonts on the system, using the
                'amssymb' package.^^J}%
       \usepackage{amssymb}%

      }{}
  \fi
\fi

\ifCUPmtlplainloaded \else
  \IfFileExists{amsbsy.sty}
    {\typeout{^^JFound the 'amsbsy' package on the system, using it.^^J}%
     \usepackage{amsbsy}}
    {\providecommand\boldsymbol[1]{\mbox{\boldmath $##1$}}}
\fi

\newsavebox{\astrutbox}
\sbox{\astrutbox}{\rule[-5pt]{0pt}{20pt}}

\title[Symmetry breaking of the reverse Bénard-von Kármán vortex street]{A model for the symmetry breaking of the reverse Bénard-von Kármán vortex street produced by a flapping foil}

\author[R. Godoy-Diana, C. Marais, J.-L. Aider and J.E. Wesfreid]{Ramiro Godoy-Diana, Catherine Marais, Jean-Luc Aider and Jos\'e Eduardo Wesfreid}
\affiliation{{Physique et M\'ecanique des Milieux H\'et\'erog\`enes (PMMH)}\\
{UMR 7636 CNRS; ESPCI; Univ. Paris 6; Univ. Paris 7}\\
{10, rue Vauquelin, F-75231 Paris Cedex 5, France}\\
}
\begin{document}

\maketitle

\begin{abstract}
The vortex streets produced by a flapping foil of span-to-chord aspect ratio of 4:1 are studied in a hydrodynamic tunnel experiment. In particular, the mechanisms giving rise to the symmetry breaking of the reverse Bénard-von Kármán vortex street that characterizes fish-like swimming and forward flapping flight are examined. Two-dimensional particle image velocimetry measurements in the mid-plane perpendicular to the span axis of the foil are used to characterize the different flow regimes. The deflection angle of the mean jet flow with respect to the horizontal observed in the average velocity field is used as a measure of the asymmetry of the vortex street. Time series of the vorticity field are used to calculate the advection velocity of the vortices with respect to the free-stream, defined as the phase velocity $U_{\mathrm{phase}}$, as well as the circulation $\Gamma$ of each vortex and the spacing $\xi$ between consecutive vortices in the near wake. The observation that the symmetry breaking results from the formation of a dipolar structure from each couple of counter-rotating vortices shed on each flapping period serves as starting point to build a model for the symmetry breaking threshold. A symmetry breaking criterion based on the relation between the phase velocity of the vortex street and an idealized self-advection velocity of two consecutive counter-rotating vortices in the near wake is established. The predicted threshold for symmetry breaking accounts well for the deflected wake regimes observed in the present experiments and may be useful to explain other experimental and numerical observations of similar deflected propulsive vortex streets reported in the literature.
\end{abstract}

\section{Introduction}
\begin{figure}
\centering
\includegraphics[width=0.7\linewidth]{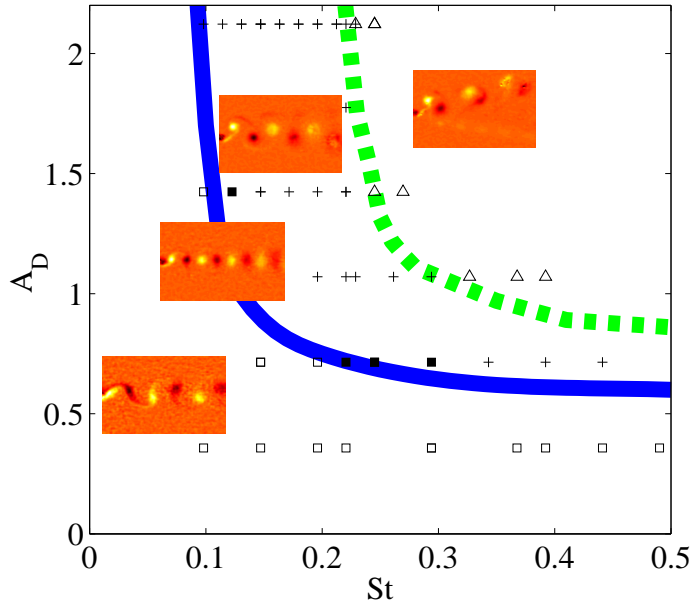}
\caption{Transitions in the wake of a flapping foil in the $A_D$ vs. $\mathrm{St}$ map for $\mathrm{Re}=255$ \cite[from][]{godoy-diana2008}. Experimental points are labeled as $\square$: BvK wake; $\blacksquare$: aligned vortices (2S wake); $+$: reverse BvK wake; $\vartriangle$: deflected reverse BvK street resulting in an asymmetric wake. Solid line: transition between BvK and reverse BvK. Dashed line: transition between reverse BvK and the asymmetric regime. Typical vorticity fields are shown as inserts on each region.}
\label{fig_phase_space}
\end{figure}

Flapping-based propulsive systems, either natural or man-made, are often discussed in terms of the Strouhal number, defined as the product of the flapping frequency $f$ and amplitude $A$ divided by the cruising speed $U_0$, i.e. $St_A = fA/U_0$ \cite[][]{anderson1998,taylor2003}. Another crucial parameter in these problems is the aspect ratio of the flapping body, because it determines to what extent a quasi-two-dimensional (Q2D) view can capture the main elements needed for an adequate description of the real three-dimensional (3D) flow. In particular, in the case of a flapping body propelling itself in forward motion, at least two qualitatively different situations have been evidenced from flapping foil experiments and numerical simulations: high span-to-chord ratio foils produce a propulsive vortex street, the reverse Bénard-von Kármán (BvK) wake \cite[see e.g.][]{koochesfahani1989,anderson1998}, where the most intense vortices are aligned with the foil span and turn in opposite senses with respect to the natural BvK vortices behind a 2D cylinder. A Q2D analysis accounts for the key dynamical features in this case where the mean flow has the form of a jet and results in a net propulsive force. As the span-to-chord ratio decreases towards unity, 3D effects come into play and modify dramatically the structure of the wake. In this case a series of vortex loops (or horse-shoe vortices) are engendered from the vorticity shed from all sides of the flapping foil \cite[see e.g.][]{vonellenrieder2003,buchholz2006,buchholz2008}. The experiments reported here were performed with a 4:1 aspect ratio foil, which is high enough to produce Q2D regimes in the near wake. A two-parameter description that permits to vary independently the frequency and amplitude of the oscillatory motion has been shown recently \citep{godoy-diana2008} to be the optimum framework to fully characterize the quasi-two-dimensional regimes observed in the wake of a pitching foil. The transition from a BvK vortex street to the reverse BvK street characteristic of propulsive regimes, and the symmetry breaking of the reverse BvK street reported in \cite{godoy-diana2008} are summarized in the $(St,A_D)$ phase space shown in figure \ref{fig_phase_space}. The Strouhal number and a dimensionless amplitude have been defined using a fixed length scale (the foil width $D$) as $St=fD/U_0$ and $A_D=A/D$, respectively \footnote{Note that the product of these two parameters gives the flapping amplitude based Strouhal number that is often used, i.e. $St\times A_D = fA/U_0 = St_A$}.

\begin{figure}
\centering
\includegraphics[width=\linewidth]{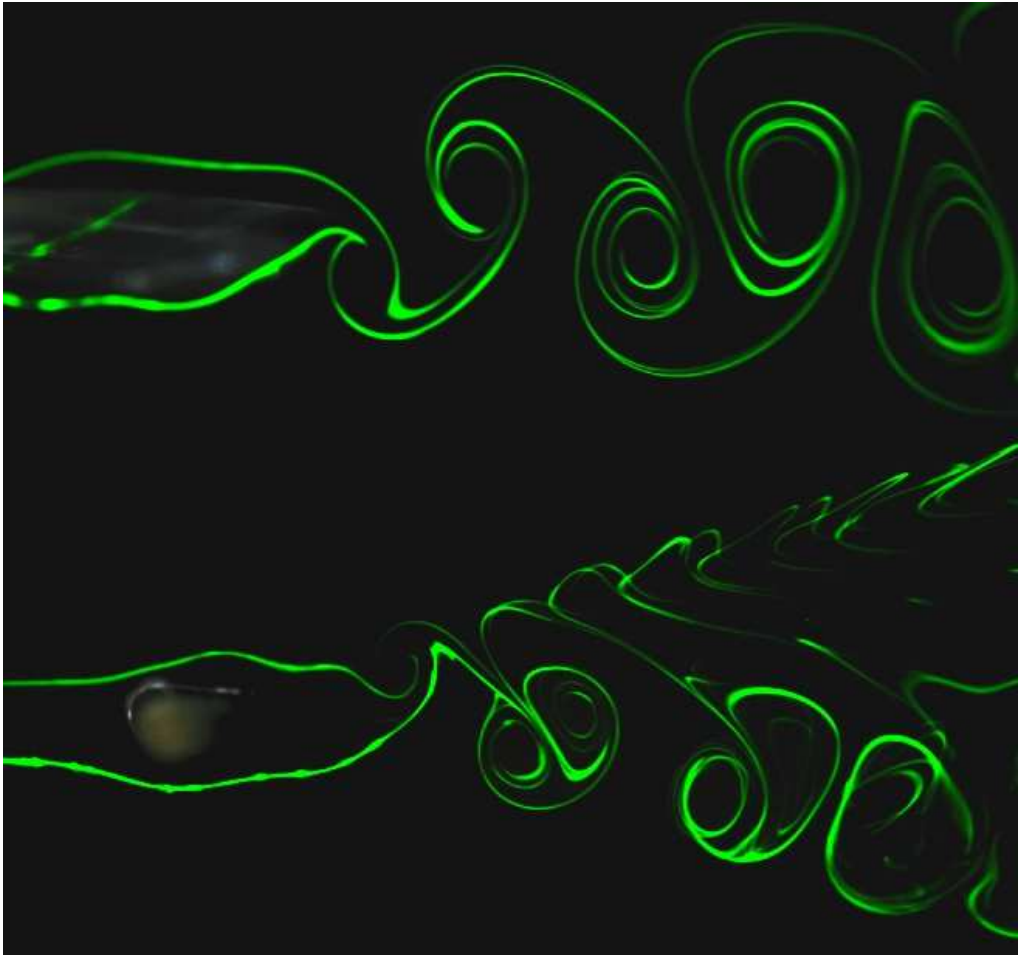} 
\caption{Fluorescein dye visualization of the typical reverse Bénard-von Kármán vortex street that characterizes flapping based propulsion (top); and an asymmetric wake (bottom) that is produced by some flapping configurations even when the flap motion is symmetric.}
\label{fluo}
\end{figure}

In this paper we focus on the reverse BvK regime, attempting to shed some light on the physical mechanisms that determine its symmetry breaking (the dashed line in figure \ref{fig_phase_space}). Asymmetric vortex streets arising in the wake of a flapping foil have been observed in many experimental \citep{jones1998,godoy-diana2008,vonellenrieder2008,buchholz2008} and numerical studies \citep{jones1998,lewin2003} even when the geometry of the problem and the flapping motion are symmetric. Other configurations of forced wakes have been shown to produce asymmetric flows that arise from the interaction of natural and forced vortex shedding ---see for instance the soap-film experiments with forced cylinder wakes in \cite{couder1986}. The deflection of the propulsive vortex street determines that the net force generated by the flapping motion is not aligned with the foil symmetry plane, or in other words that a mean lift force accompanies the production of thrust. A qualitative explanation of the process that determines the symmetry breaking can be given analyzing the structure of the wake: the deflection of the vortex street that signals the symmetry breaking results from the formation of a dipolar structure from each couple of counter-rotating vortices shed on each flapping period (see figure \ref{fluo}). Above a certain threshold, the self-advection of the dipolar structure formed over one flapping period is strong enough to decouple from the subsequent vortex in the street and to generate a deflection of the mean flow. These asymmetric wakes occur in a region of the parameter space that overlaps the high-efficiency Strouhal number range used by flapping animals, which makes the precise definition of the symmetry breaking threshold potentially important for the design of artificial flapping-based propulsors and their control. This is the goal of the present work, where we study in detail the spanwise vorticity field in the near wake of the flapping foil in order to characterize all the basic features of the reverse BvK vortex streets.  Relying on a hypothesis of quasi-two-dimensionality of the flow in the near wake, we propose a predictive symmetry breaking criterion based on the phase velocity of the vortex street.

\begin{figure}
\centering
\includegraphics[width=0.8\linewidth]{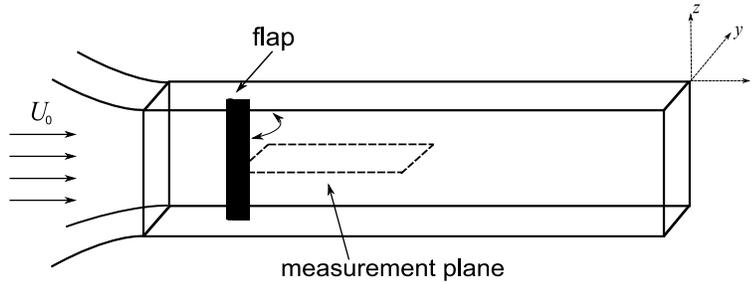}
\caption{Schematic diagram of the experimental setup in the hydrodynamic tunnel.}
\label{fig_exp_setup}
\end{figure}
\section{Experimental setup}
The setup is the same described in \cite{godoy-diana2008} and consists of a pitching foil placed in a hydrodynamic tunnel (see figure \ref{fig_exp_setup}). The foil chord $c$ is 23mm and its span is 100mm which covers the whole height of the 100 $\times$ 150mm section of the tunnel. The foil profile is symmetric, opening at the leading edge as a semicircle of diameter $D=5$mm which is also the maximum foil width. The pitching axis is driven by a stepper motor. The control parameters are the flow velocity in the tunnel $U_0$, the foil oscillation frequency $f$ and peak-to-peak amplitude $A$ which let us define the Reynolds number, the Strouhal number and a dimensionless flapping amplitude as, respectively:

\begin{equation}
  Re=UD/\nu\;,\;\; St=fD/U_0\;,\;\; A_D=A/D \;.
\end{equation}

\noindent where $\nu$ is the kinematic viscosity. In the strongly forced regimes produced by the flapping foil, the flapping frequency used to define $St$ is equivalent to the main vortex shedding frequency. The boundary layer thickness on the tunnel walls at the position of the flap is of approximately 10mm for the present experiments. Measurements were performed using 2D Particle Image Velocimetry (PIV) on the horizontal mid-plane of the flap. PIV acquisition and post-processing was done using a LaVision system with an ImagerPro $1600 \times 1200$ 12-bit CCD camera recording pairs of images at $\sim{15}$Hz and a two rod Nd:YAG (15mJ) pulsed laser. Laser sheet width was about 1mm in the whole 100mm $\times$ 80mm imaging region. The time lapse between the two frames (d$t$) was set to 12ms. Additional post-processing and analysis were done using Matlab and the PIVMat Toolbox \citep{pivmat}.

\begin{figure}
\centering
\includegraphics[height=0.65\linewidth]{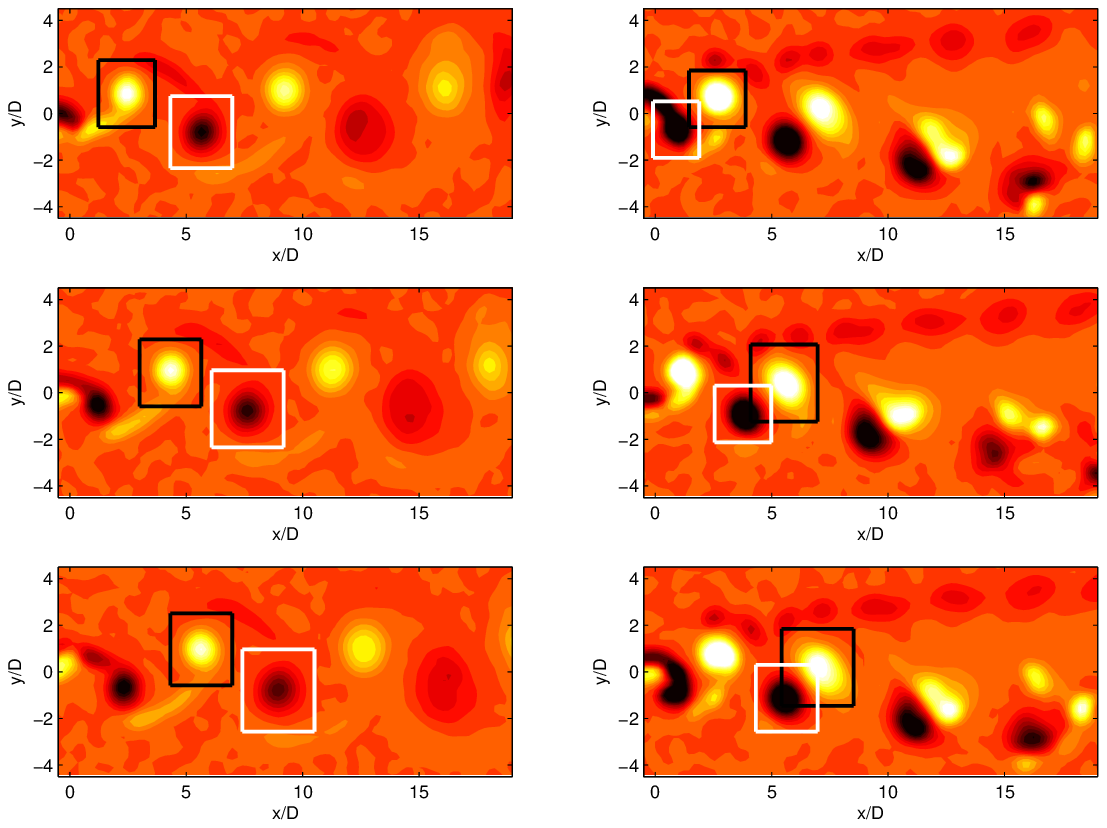}\includegraphics[height=0.65\linewidth]{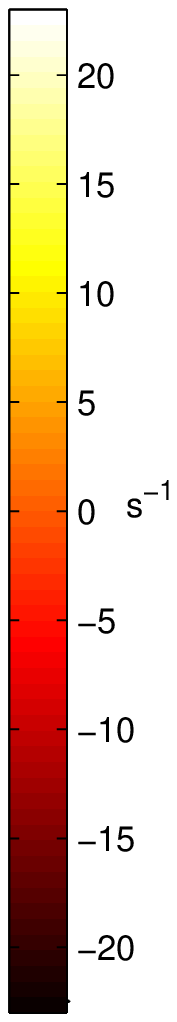}
\caption{Three consecutive snapshots of the vorticity field in the wake of the flapping foil at $Re=255$ and $A_D=1.42$ for two different flapping frequencies. Left column: $St=0.15$, reverse BvK wake; and right column: $St=0.27$, asymmetric wake.}
\label{fig_vorticity_sequence}
\end{figure}

\begin{figure}
\centering
\includegraphics[width=1\linewidth,trim=0 0 0 0,clip=true]{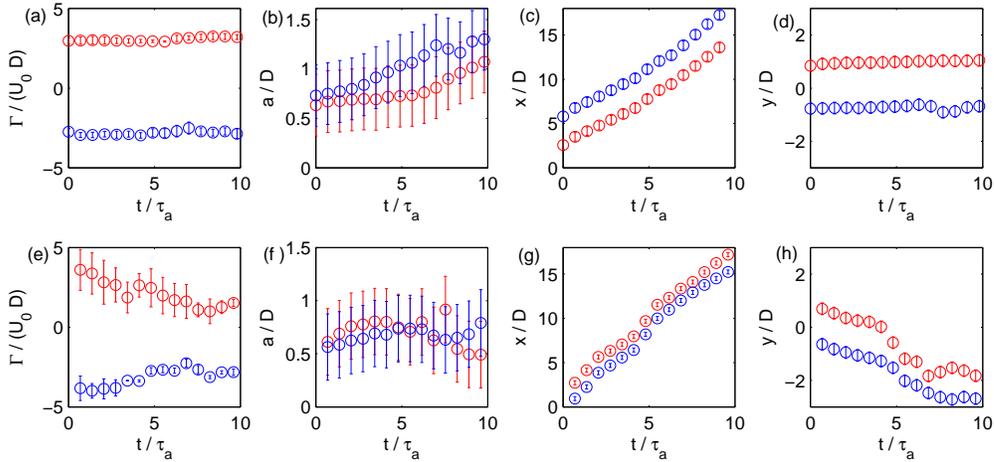}
\caption{Circulation (a,e), radius (b,f), $x$-position (c,g) and $y$-position (d,h) of two consecutive (and counter-rotating) vortices in the wake of the flapping foil as a function of time.  The top figures (a-d) correspond to the reverse BvK wake and the bottom figures (e-h) to the asymmetric wake shown in figure \ref{fig_vorticity_sequence} (left and right columns, respectively).}
\label{fig_circulation_etal}
\end{figure}

\section{Observations}

\subsection{The vorticity field}

The $\omega_z$ vorticity field is calculated from the $u_x$ and $u_y$ fields obtained from PIV using 2nd-order centered differences. Two sequences of vorticity fields are shown in figure \ref{fig_vorticity_sequence}: the left column corresponds to a reverse BvK street and the right column to an asymmetric wake. The Reynolds number and flapping amplitude are the same in both cases and only the Strouhal number has been increased from 0.15 for the reverse BvK to 0.27 for the asymmetric wake. The positions of two consecutive (and counter-rotating) vortices are followed for each experiment using a search of local maxima $(X_{max}(t),Y_{max}(t))$ and minima $(X_{min}(t),Y_{min}(t))$ in the $\omega_z$ field. These are the coordinates of the center of the rectangles in figure \ref{fig_vorticity_sequence}.

\begin{figure}
\centering
\includegraphics[width=0.8\linewidth]{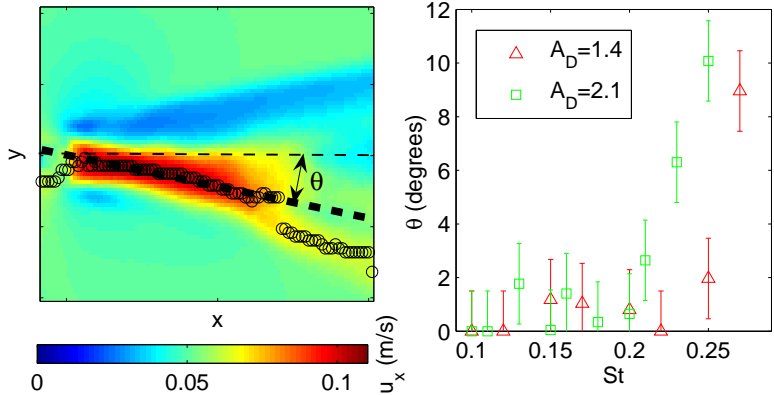}
\caption{Definition of the asymmetry angle $\theta$ from the deflected mean jet flow (left) and $\theta$ as a function of the Strouhal number for $A_D=1.4$ and $2.1$ (right). The circles shown in the left plot are the maxima of cross-stream profiles of the average horizontal velocity field. A linear fit over these points in the near wake gives the dotted line that defines the angle theta. The error bars in the right plot represent a constant $\pm1.5$ degrees that was the maximum deflection angle measured with the previous method for a wake in the symmetric reverse Bénard-von Kármán configuration.}
\label{fig_asymmetry}
\end{figure}

\subsection{The circulation $\Gamma$}
An area of integration that encompasses each vortex needs to be defined in order to calculate its characteristic features such as circulation and size. We use gaussian fits $~\exp(-x_i^2/\sigma_i^2)$ along the vertical and horizontal axes centered on the positions of the maxima and minima of vorticity and define the sizes of the vortex along each direction $x$ and $y$ as $2\sigma_i$. These are the horizontal and vertical sizes of the rectangles in the time sequences of figure \ref{fig_vorticity_sequence}. The choice of rectangular integration contours (instead of the elliptical ones that would have better followed the vortex shape) was kept in order to avoid introducing error from interpolation of the PIV data. Since the vortex cores are nearly circular we define a single vortex radius $a$ as the mean of the sizes calculated along the two principal axes. The circulation $\Gamma$ can be then calculated either from a line integral of the velocity field on each contour or a surface integral of the vorticity field over each rectangle. Although in theory the two definitions of $\Gamma$ are equivalent, the two calculations from the experimental fields are slightly different mainly because, vortices in the wake being not too far from each other, the contour for the line integral calculation sometimes passes through a neighboring opposite-signed vortex (see figure \ref{fig_vorticity_sequence}, right column) and gives thus a spurious contribution to the total circulation. The circulation plotted in figure \ref{fig_circulation_etal} is the mean value between the two different methods of calculation, the difference giving the error bars. When the integration regions for the two counter-rotating vortices do not overlap, as is the case in the left column of figure \ref{fig_vorticity_sequence}, the difference in the two calculations is small (see figure \ref{fig_circulation_etal}.a). On the contrary, the error bars are larger when the two rectangles overlap, as in the case of the asymmetric wake depicted in the right column of figure \ref{fig_vorticity_sequence}. This can be seen in figure \ref{fig_circulation_etal}.e, where the spurious effect due to having the neighboring vortex partially overlapping the integration region appears as a decreasing trend (in absolute value) in the time evolution of  $\Gamma$. We also plot in figure \ref{fig_circulation_etal} the time evolution of the vortices radii and positions. It can be seen in figure \ref{fig_circulation_etal}.h that the deflection of the wake is correctly captured by the time evolution of the $y$-coordinate. The time axes in all these plots are nondimensionalized by the advection time scale $\tau=D/U_0$.

\subsection{The asymmetric wake}

\begin{figure}
\centering
\includegraphics[width=0.85\linewidth]{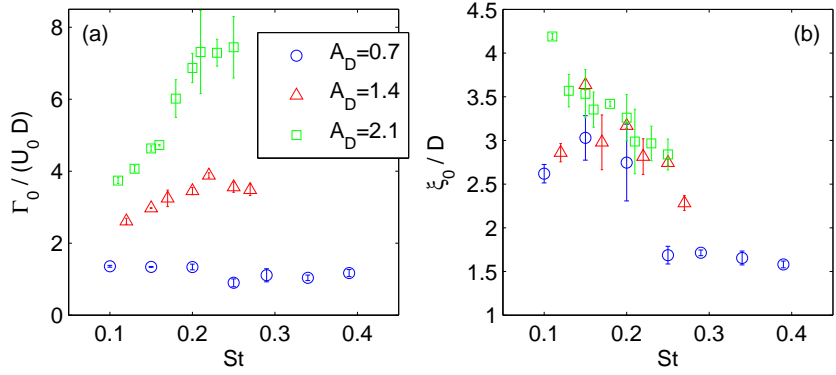}
\caption{(a) Circulation $\Gamma_0$ and (b) distance between two consecutive vortices $\xi_0$ as a function of $St$ and $A_D$. Circles, triangles and squares correspond, respectively, to $A_D=0.7, 1.4$ and $2.1$.}
\label{fig_circ_dist_St_AD}
\end{figure}

The domain of existence of the reverse BvK vortex street is bounded on the upper-right zone of the $(St,A_D)$ phase space by a transition to an asymmetric regime (figure \ref{fig_phase_space}). In order to characterize this transition we define an angle of asymmetry $\theta$ using the direction of the jet observed in the mean velocity field (see figure \ref{fig_asymmetry}). Of the three series with different flapping amplitudes studied here (see figure \ref{fig_circ_dist_St_AD} in the next section), the smallest one ($A_D=0.7$) does not produce an asymmetric wake so only the cases of $A_D=1.4$ and $2.1$ are reported in figure \ref{fig_asymmetry}. These measurements show that the transition is rather abrupt and support the idea of the existence of a symmetry breaking threshold.

\section{Discussion}

\subsection{Back to the $(St,A_D)$ phase space}
Thinking in a $(St,A_D)$ phase space perspective, we use the 'initial' values of the circulation and the positions of the maxima and minima of $\omega_z$, defined as the values measured when the positive vortex crosses a reference frame fixed at $X_{\mathrm{ref}}=2.5D$, in order to compare the vortices produced by different flapping configurations. We thus plot in figure \ref{fig_circ_dist_St_AD} the initial value of the circulation $\Gamma_0$ and the distance between the two consecutive counter-rotating vortices $\xi_0$ , as a function of $St$ for three series with different values of $A_D$. The definition of $\xi$ in terms of the positions of two counter-rotating vortices shown in figure \ref{fig_circulation_etal} is, as shown graphically in figure \ref{fig_model_v_phase_eff} (left), $\xi=\sqrt{(X_{max}-X_{min})^2+(Y_{max}-Y_{min})^2}$. The main observation is that, for a given amplitude, the circulation increases and the distance between consecutive vortices decreases with increasing flapping frequency. Although this is not surprising, it is an important point because it lets us ascertain that a threshold curve determining the symmetry breaking of the reverse BvK wake can be traced in a $(\Gamma_0,\xi_0)$ space and also that a model containing the basic physics of the problem could be tested using experimental measurements.

\subsection{Phase velocity}

\begin{figure}
\centering
\includegraphics[width=0.87\linewidth]{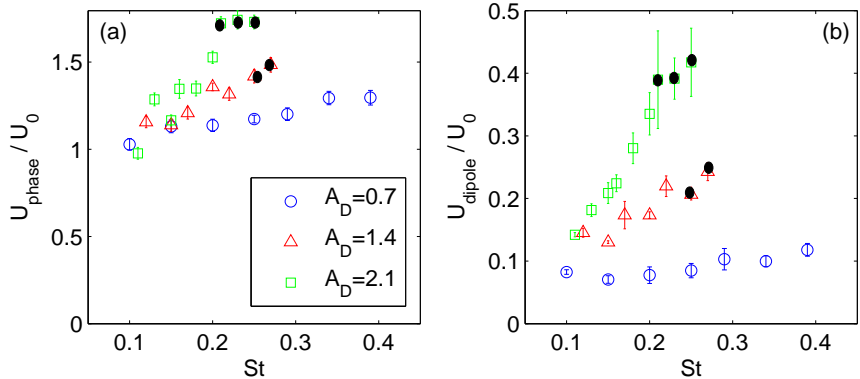}
\caption{(a) Phase velocity $U_{\mathrm{phase}}$ of the vortex street normalized with respect to the free-stream $U_0$; and (b) translation speed of the initial dipoles $U_{\mathrm{dipole}}$ calculated using Eq. \ref{v_dipole_formula}. Both quantities are plotted as a function of $St$ and $A_D$. Circles, triangles and squares correspond, respectively, to $A_D=0.7, 1.4$ and $2.1$. Asymmetric wakes are marked with a dark dot.}
\label{fig_u_phase_u_dipole}
\end{figure}

The advection velocity of the vortices with respect to the free-stream can be calculated from the $x$-position measurements $X_i(t)$ in figures \ref{fig_circulation_etal}.c and \ref{fig_circulation_etal}.g. We define this \emph{phase speed} $U_{\mathrm{phase}}$ as the slope $dX_i/dt$ evaluated at the initial reference time \footnote{Vortices slightly accelerate in the initial part of the wake so that the wavelength measured in a snapshot of the vorticity field over the vortices nearest to the flap is shorter than the wavelength measured farther in the wake \cite[see also][]{bearman1967}. Here we deliberately define $U_{\mathrm{phase}}$ as the advection speed at the initial reference time because we are interested in the near wake mechanism that triggers the symmetry breaking. }.
In the case of the Bénard-von Kármán vortex street past a bluff body $U_{\mathrm{phase}}<U_0$ \cite[see e.g.][]{bearman1967,williamson1989}, which is related to the velocity deficit in the wake of the obstacle. The phase velocity for the present experiments is plotted in figure \ref{fig_u_phase_u_dipole} (left). The propulsive nature of the reverse BvK wake appears clearly in the fact that $U_{\mathrm{phase}}>U_0$.

\subsection{The dipole model}

The physical mechanism giving rise to a deflected wake is based on the formation of a dipolar structure on each flapping period \citep{godoy-diana2008}, a feature that has also been observed in forced wakes in soap films \cite[][]{couder1986}, supporting the idea of a mainly 2D phenomenon. The initial condition sets the choice of the side where the asymmetry develops: the first dipole that is formed entrains fluid behind it, deflecting the mean flow in the wake and forcing the subsequent dipoles to follow the same path. This initial perturbation exists for all the $(St,A_D)$ parameter space, however, after a few periods of symmetric flapping motion, only in a region above a certain threshold (see figure \ref{fig_phase_space}) the wakes remain asymmetric. The two quantities $(\Gamma_0,\xi_0)$ plotted in figure \ref{fig_circ_dist_St_AD} can be used to give a measure of the strength of the dipolar structures that are formed for each flapping configuration. The simplest model that contains $\Gamma$ and $\xi$ is to consider a dipole made of two point vortices of circulations $\pm\Gamma$ separated by a distance $\xi$. In this case, the translation speed of the dipole (determined by the effect of each vortex over the other) is given by \citep[see e.g.][]{saffman}

\begin{equation}\label{v_dipole_formula}
U_{\mathrm{dipole}}=\frac{\Gamma}{2 \pi \xi} \; .
\end{equation}

The values of $U_{\mathrm{dipole}}$ calculated using the data from figure \ref{fig_circ_dist_St_AD} are shown in figure \ref{fig_u_phase_u_dipole} (right). As a direct consequence of the behavior of $\Gamma_0$ and $\xi_0$ in figure \ref{fig_circ_dist_St_AD}, for each flapping amplitude, the self-induced speed of the dipolar structure increases with the Strouhal number. It is remarkable that $U_{\mathrm{dipole}}$ can reach values up to almost $50\%$ of $U_0$. The two plots in figure \ref{fig_u_phase_u_dipole} show that a correlation exists between $U_{\mathrm{dipole}}$ and $U_{\mathrm{phase}}$. This can be easily understood considering that the flow field produced by each vortex, and hence $U_{\mathrm{dipole}}$, contributes to the overall advection velocity in the vortex street.

\begin{figure}
\centering
\includegraphics[width=0.45\linewidth]{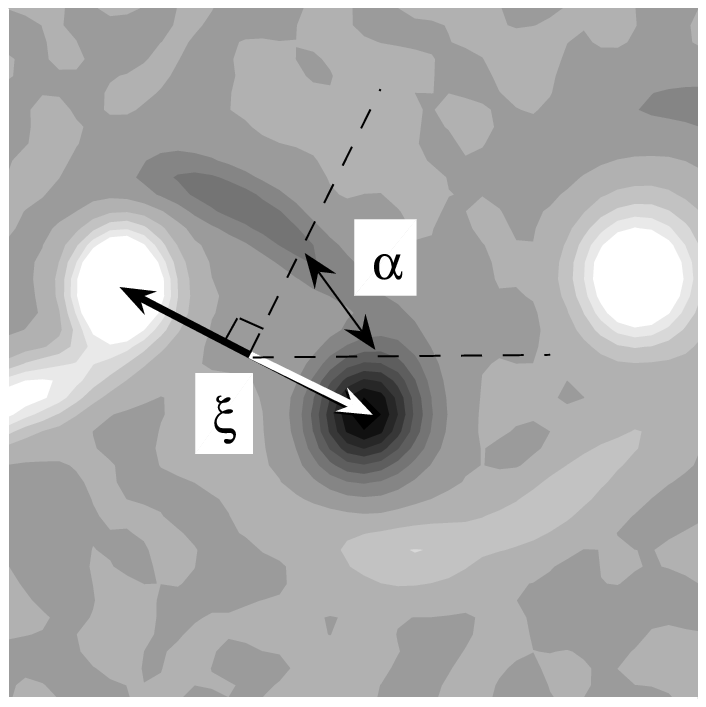}\includegraphics[width=0.44\linewidth]{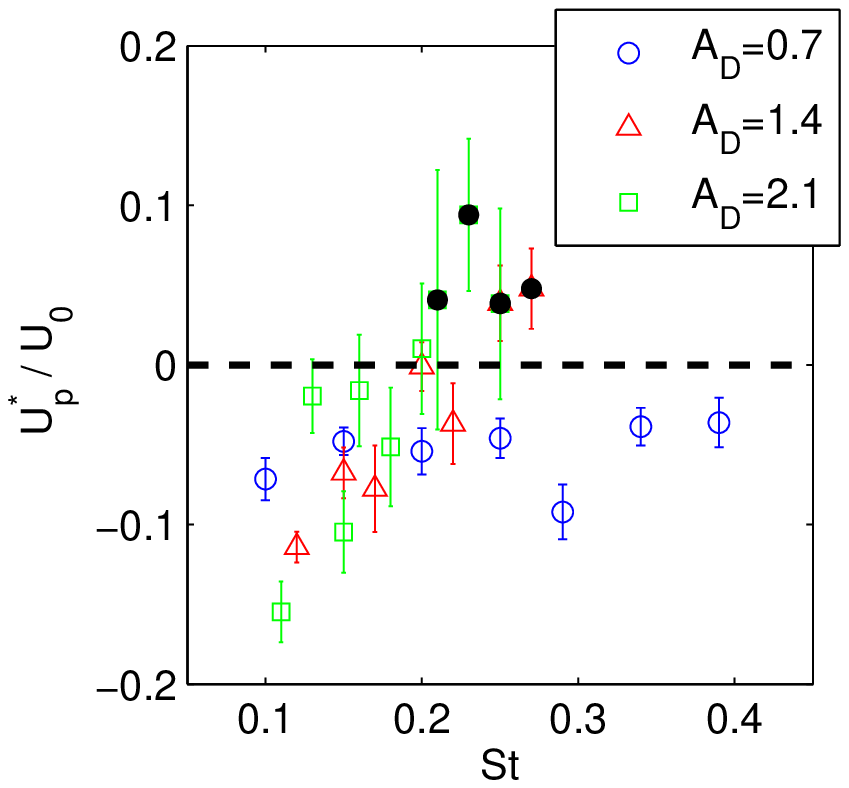}
\caption{Left: Definition of the angle $\alpha$ used in equation \ref{v_phase_eff}. The line at angle $\alpha$ with the horizontal is at right angles with the segment of length $\xi$ that joins the two consecutive counter-rotating vortices. Right: Effective phase velocity $U_{\mathrm{p}^*}$ normalized with respect to the free-stream $U_0$ plotted as a function of $St$ and $A_D$. Circles, triangles and squares correspond, respectively, to $A_D=0.7, 1.4$ and $2.1$. Asymmetric wakes are marked with a dark dot. The points above the dashed axis verify the proposed symmetry breaking condition $U_{\mathrm{p}}^*>0$.}
\label{fig_model_v_phase_eff}
\end{figure}

Thinking about the definition of a criterion for the symmetry breaking, it seems evident from figure \ref{fig_u_phase_u_dipole} that no single threshold can be established for $U_{\mathrm{dipole}}$ or $U_{\mathrm{phase}}$ that accounts for all observations of asymmetric wakes (we have marked these in figure \ref{fig_u_phase_u_dipole} with black dots). Because we have observed that the symmetry breaking is related to the ability of a given dipolar structure to escape from the ``symmetrizing" effect of the subsequent vortices in the wake, it is useful to find a quantitative measurement of the velocity at which vortices are moving in the direction defined by the dipole. In order to do so we define the \emph{effective phase velocity}, in terms of the angle $\alpha$ between the horizontal (the streamwise direction) and the direction of $U_{\mathrm{dipole}}$ (see figure \ref{fig_model_v_phase_eff}), as:

\begin{equation}\label{v_phase_eff}
U_{\mathrm{p}}^*= (U_{\mathrm{phase}}-U_0)\cos\alpha-U_{\mathrm{dipole}} \; .
\end{equation}

\noindent Recalling that the measured phase velocity $U_{\mathrm{phase}}$ results from the superposition of the free-stream velocity and all the velocities induced by the vortices in the wake, the effective phase velocity defined in equation \ref{v_phase_eff} is actually the component along the dipole direction of the velocity induced by the vortex street excepting the contribution of the dipole being considered, and not including the free-stream velocity. Within the limits imposed by the experimental uncertainty, it can be seen in figure \ref{fig_model_v_phase_eff} that a fair prediction of the observed behavior can be given saying that the reverse BvK vortex street will be prone to breaking symmetry if the effective phase velocity $U_{\mathrm{p}}^*$ is positive, which is when the vortex street downstream of the dipole being considered enhances the dipole velocity.

\section{Concluding remarks}

Reverse Bénard-von Kármán (BvK) vortex streets are a fundamental feature of fish-like swimming and forward flapping flight and their symmetry properties are intimately related to the cycle of thrust and lift production. Although the deflection of these propulsive wakes has been observed and characterized in various flapping-foil experiments \citep{jones1998,godoy-diana2008,vonellenrieder2008} and numerical simulations \citep{lewin2003}, the model proposed in the present paper is, to our knowledge, the first attempt to produce a quantitative threshold prediction based on the observed physical mechanism underlying the symmetry breaking. In spite of the strongly idealized dipolar model used in Eq. \ref{v_dipole_formula} as a fundamental element of the model, the predicted threshold accounts reasonably well for the experimental reality and should be useful to rationalize the similar observations reported in the literature.

\thanks{We gratefully acknowledge Maurice Rossi for fruitful discussions on the dipole model.}

\end{document}